\documentclass[12pt]{article}

\pdfoutput=1 

\usepackage{amsmath, amssymb, amsthm, amsbsy, bbm}

\usepackage{graphics, color}
\usepackage{graphicx}
\usepackage[normalem]{ulem}
\usepackage{cite}
\usepackage[colorlinks=true,urlcolor=blue,anchorcolor=blue,citecolor=blue,filecolor=blue,linkcolor=blue,menucolor=blue,linktocpage=true,pdfproducer=medialab,pdfa=true]{hyperref}

\newcommand{\sect}[1]{\section{#1}\setcounter{equation}{0}}

\newcommand{\be}{\begin{equation}}
\newcommand{\ee}{\end{equation}}
\newcommand{\OL}[1]{\,\overline{\! #1 \!}\,}

\begin{document}


\begin{titlepage}
\bigskip
\bigskip\bigskip\bigskip
\centerline{\Large  Models of AdS$_2$ Backreaction and Holography}
\bigskip\bigskip\bigskip
 \centerline{{\bf Ahmed Almheiri}\footnote{\tt almheiri@stanford.edu}}
\medskip
\centerline{\em Stanford Institute for Theoretical Physics}
\centerline{\em Department of Physics, Stanford University}
\centerline{\em Stanford, CA 94305}
\bigskip 
\centerline{\em Department of Physics}
\centerline{\em University of California}
\centerline{\em Santa Barbara, CA 93106}\bigskip
\bigskip
 \centerline{{\bf Joseph Polchinski}\footnote{\tt joep@kitp.ucsb.edu}}
\medskip
\centerline{\em Kavli Institute for Theoretical Physics}
\centerline{\em University of California}
\centerline{\em Santa Barbara, CA 93106-4030}\bigskip
\bigskip
\bigskip\bigskip

\begin{abstract}
We develop models of 1+1 dimensional dilaton gravity describing flows to $AdS_2$ from higher dimensional $AdS$ and other spaces.  We use these to study the effects of backreaction on holographic correlators.  We show that this scales as a relevant effect at low energies, for compact transverse spaces.  We also discuss effects of matter loops, as in the CGHS model.
\end{abstract}
\end{titlepage}

\baselineskip = 16.7pt

\tableofcontents

\setcounter{footnote}{0}

\bigskip
\bigskip
\sect{Introduction}

Many holographic theories flow to an $AdS_2 \times X$ geometry in the infrared.  For example, this is true for Reissner-Norstrom black holes~\cite{BR}, and correspondingly for a large class of finite density systems~\cite{Chamblin:1999tk} (see~\cite{Iqbal:2011in} for a recent review of applications).  The $AdS_2/CFT_1$ system is rather exotic, in that only the time coordinate transforms under scaling.  The case that $X$ is compact is particularly problematic, for a number of reasons.  One is that the backreaction is so strong that the theory has no excitations~\cite{Maldacena:1998uz}.  This raises a puzzle, in that holographic calculations of correlators seem to give typical conformal behaviors $(t-t')^{-2\Delta}$: how can there be nontrivial time-dependence in a system that has no finite energy states?

In order to investigate this we develop a class of toy models in which the backreaction problem can be studied.  These are similar in spirit to the CGHS model~\cite{Callan:1992rs}.  They are inspired by the dimensional reduction of the magnetic $AdS_2$ vacua studied in Refs.~\cite{Hartnoll:2007ai,D'Hoker:2009mm,Almheiri:2011cb}, which flows from $AdS_4$ in the UV to $AdS_2$ in the \mbox{IR}.  
In Sec.~2 we introduce the models and study some general features, including their static solutions and asymptotic behaviors.  In Sec.~3 we focus on a special case which is simple enough to solve completely at the classical level.  This model flows from a conformal Lifshitz behavior in the UV to $AdS_2$ in the IR.  The UV geometry regulates the backreaction and allows finite energy states.  We then study the response of the geometry to an infalling matter pulse.  In Sec.~4 we use this system as a toy model of holography, calculating the 2- and 4-point functions at leading order in $1/N$.  We find that the IR behavior of the 4-point function is not conformal, but actually relevant: the backreaction makes this symmetry anomalous, at least in its action on finite energy states.  Thus, for compact $X$ it appears that the conformal low energy sector consists only of the zero energy states, as emphasized in Ref.~\cite{Sen:2008vm}.\footnote{Compact $X$ also brings in the possibility of $AdS_2$ fragmentation~\cite{Maldacena:1998uz}.  This is usually an issue for $X=S^2$; we are imagining that $X = T^2$, as would arise from compactification of a condensed matter system.  Note that the fragmentation geometry is intrinsically four-dimensional and would not be seen in a two-dimensional reduction.}   In Sec.~5 we include quantum effects in the large-$N$ approximation as is done in the CGHS model~\cite{Callan:1992rs}.  We discuss puzzles regarding the density of states.


\sect{AdS$_2$ back-reaction models}

\subsection{The models}

We will consider the family of 1+1 dimensional models
\begin{equation}
L = {1 \over 16\pi G}  \sqrt{-g}\left\{ \Phi^2 R + \lambda (\nabla \Phi)^2 - U(\Phi)   \right\} \,, \label{L1}
\end{equation}
characterized by the parameter $\lambda$ and the potential $U(\Phi)$.   Here $1/G$ is proportional to $V$, the volume of $X$.  More general models of dilaton gravity are reviewed in Refs.~\cite{deAlwis:1992hv,Grumiller:2002nm}.  The Weyl transformation 
\be
g_{a b} \rightarrow g_{a b} \Phi^{-\alpha /2} \label{weyl}
\ee
shifts
\begin{equation}
\lambda \to \lambda - \alpha\,,\quad U(\Phi) \to \Phi^{-\alpha/2} U(\Phi) \,.
\end{equation}
Thus $\lambda$ can be set to zero without loss of generality; we denote the resulting potential by $ U_0(\Phi)= \Phi^{-\lambda/2} U(\Phi)$.  The field $\Phi$ will still have a kinetic term from $\Phi^2 R$.  Note that we are neglecting a possible anomaly in the Weyl transformation, as appropriate for the classical considerations of \S2-4.  In \S5 we will return to this issue.

Some examples are
\begin{itemize}
\item The CGHS model~\cite{Callan:1992rs}: $\lambda = 4$, $U(\Phi) = - A \Phi^2$, $A>0$.  This is obtained via dimensional reduction in the throat limit of near extremal dilatonic black holes in four or five dimensions~\cite{Giddings:1992kn}.
\item Magnetic branes~\cite{Hartnoll:2007ai,D'Hoker:2009mm,Almheiri:2011cb}: $\lambda = 2$, $U(\Phi) =  B^2/\Phi^2 - A \Phi^2$, $A>0$.  This system arises by turning on a Kaluza-Klein magnetic field in the near-horizon $N$-M2 brane geometry, with the possibility of an additional $Z_k$ orbifolding.   This admits a rich phase structure as a function of $N$ and $k$.   The geometry interpolates from $AdS_4$ in the UV and AdS$_2 \times R^2$ in the IR. 
The two-dimensional model is obtained by reduction of the metric
\begin{equation}
ds_4^2 = g_{\mu\nu}dx^\mu dx^\nu + \Phi^2(x)(dy_1^2 + dy_2^2) \,. \label{ansatz}
\end{equation}
\item A toy model: $\lambda = 0$, $U(\Phi) = C - A \Phi^2 $, $A>0$, $C>0$.  This does not arise from any particular reduction, but has the convenient properties that its dynamics is classically solvable and it has a solution that interpolates between a conformal Lifshitz spacetime in the UV and AdS$_2$ in the IR. 
\end{itemize}

As in the CGHS model, we will also add matter fields $f$,
\begin{align}
L_f &= {1 \over 32\pi G} \sqrt{-g} \ {\Omega(\Phi) } (\nabla f)^2 \,. \label{Om}
\end{align}
For fields moving freely in the higher dimensional spacetime, the factor $\Omega(\Phi)$ arises from the volume of the transverse dimensions.  For simplicity we will focus on models with $\Omega  = 1$.\footnote{These could arise from fields localized on defect branes~\cite{Kachru:2009xf,Jensen:2011su}, or in models where the dilaton is the string dilaton and the scalars are RR excitations~\cite{Callan:1992rs}. In any event, we expect that the inclusion of the dilaton in the scalar field would not modify the results significantly, as the $AdS_2$ dynamics is deep in the IR region where the dilaton approaches a constant.}
For now we take a single matter field, but will introduce a large number in \S5 to control quantum corrections.

\subsection{Conformal gauge}

For the most part, we will work in conformal gauge, 
\begin{equation}
ds^2 = -e^{2 \omega(x^+,x^-)}dx^+ dx^- \,,\label{conformal}
\end{equation}
with $x^\pm = t \pm z$.
The action, transformed to $\lambda = 0$ and including boundary terms, is
\begin{align}
S &= {1 \over 16 \pi G} \int d^2x \sqrt{-g}\left( \Phi^2 R -  U_{0}(\Phi) - {\Omega(\Phi) \over 2 } (\nabla f)^2  \right)  + {1 \over 8 \pi G} \int dt \sqrt{-\gamma} \Phi^2  K \\
&= {1 \over 8 \pi G} \int dt\, dz \left( \Phi^2 (\partial_t^2 - \partial_z^2) \omega -  {e^{2 \omega} \over 2}U_{0}(\Phi)  - {1 \over 4} f \partial_t(\Omega  \partial_t f) + {1 \over 4} f \partial_z( \Omega  \partial_z f)   \right)  \nonumber \\
&\qquad
 + {1 \over 8 \pi G}  \int dt \left( - \Phi^2 \partial_z \omega + {\Omega \over 4} f \partial_z f \right) \,.
\label{fullact}
\end{align}
The normalization factor, proportional to the volume of $X$, enters in the quantum discussion of \S5.
The equations of motion are
\begin{align}
2 \Omega^{-1} \partial_+ (\Omega\partial_-) f +  2 \Omega^{-1} \partial_-(\Omega\partial_+) f &= \partial_t^2 f - \partial_z^2 f 
- \partial_z \Omega \partial_z f = 0 \,, \label{gef}\\
4 \partial_+ \partial_- \Phi^2 - e^{2 \omega } U_{0}(\Phi) &= 0 \,, \label{gephi}\\
4 \partial_+ \partial_- \omega   - {e^{2 \omega } \over 2} \partial_{\Phi^2} U_{0}(\Phi) &= \partial_{\Phi^2} \Omega(\Phi) \partial_+ f \partial_-f \,, \\
-e^{2 \omega } \partial_+ \left( e^{-2 \omega } \partial_+ \Phi^2 \right) &= {\Omega \over 2} \partial_+ f \partial_+ f \,, \label{gC1}\\
-e^{2 \omega } \partial_- \left( e^{-2 \omega } \partial_- \Phi^2 \right) &= {\Omega \over 2} \partial_- f \partial_- f \label{gC2} \,.
\end{align}

\subsection{Static vacuum solutions}

We now consider static solutions, depending only on $z$, with $f=0$.  The equations of motion become
\begin{align}
(\Phi^2)'' + e^{2 \omega } U_{0}(\Phi) &= 0 \,, \label{stat1}\\
2\omega''   + e^{2 \omega }  \partial_{\Phi^2} U_{0}(\Phi) &= 0\,,  \label{stat2} \\
 \left( e^{-2 \omega } (\Phi^2)' \right)' &= 0 \label{stat3} \,.
\end{align}

Let us first consider the special case that $\Phi(z) = \OL\Phi$ is constant.  Eq.~(\ref{stat1}) requires that  $U_0(\OL\Phi) = 0$, while Eq.~(\ref{stat2}) becomes
\begin{equation}
2 \omega''   = - e^{2 \omega }  \partial_{\Phi^2} U_{0}(\OL\Phi) \,.
\end{equation}
That is, the metric is of constant curvature.  For $\partial_{\Phi^2} U_{0}(\OL\Phi) = -2 /R^2 < 0$, the curvature is negative and there are three static solutions
\be
e^{2\omega} = \frac{R^2}{z^2} \,,\quad \frac{R^2}{\sinh^2 z} \,, \quad \frac{R^2}{\sin^2 z} \,. \label{statmets}
\ee
These are respectively the Poincar\'e patch of $AdS_2$, an $AdS_2$ black hole with horizon at $z = \infty$ (or equivalently a Rindler subspace of the Poincar\'e patch), and global $AdS_2$.  For $\partial_{\Phi^2} U_{0}(\OL\Phi) = 2 /L^2 < 0$,
\be
e^{2\omega} = \frac{L^2}{\cosh^2 z}
\ee
is the static patch of $dS_2$.

For $\Phi'$ not identically zero, we can integrate Eq.~(\ref{stat3}),
\be
(\Phi^2)' = -c_1 e^{2\omega}
\ee
with nonzero $c_1$.  Defining a prepotential $U_{0}(\Phi) = \partial_{\Phi^2} W(\Phi)$, Eq.~(\ref{stat1}) then becomes
\be
\left( -c_1 (\Phi^2)' + W(\Phi)\right)' = 0 \ \Rightarrow\             dz = c_1 \frac{d(\Phi^2)}{W(\Phi) - c_2} \,.
\label{dz}
\ee
 Eq.~(\ref{stat2}) is then identically satisfied.  
 
For the magnetic brane, $W= -2B^2/\Phi - 2 A \Phi^3/3$.  At large $\Phi$ the integral of the RHS of~(\ref{dz}) converges, giving a boundary at a finite point that we take to be $z=0$, with the asymptotic behavior
\be
\Phi^2 \propto 1/ z^2\,, \quad e^{2\omega} \propto 1/ z^3 \,.
\ee
The lift~(\ref{ansatz}), including the shift back to $e^{2\omega({\lambda = 2})} = e^{2\omega}/\Phi$,
  gives an $AdS_4$ geometry.  The prepotential $W$ has a maximum  $\OL W = -8 A^{1/4} |B|^{-3/2}/3$ at $\OL\Phi= |B|^{1/2} A^{-1/4}$.  For $c_2 = \OL W$, the solution flows from $AdS_4$ to the $AdS_2$ solution described above at large $z$ (times a $T^2$ from the reduction).  For $c_2 > \OL W$, $\Phi$ goes to zero at a finite value of $z$, producing a naked singularity.  For $c_2 < \OL W$, $z$ diverges logarithmically as the zero of the denominator is approached, and the solution is a black hole.  

For the toy model the prepotential is $W =  C\Phi^2 - A \Phi^4/2$.  Again Eq.~(20) is integrable at large $\Phi$, giving the $z \to 0$ behavior
\be
\Phi^2 \propto 1/z\,,\quad e^{2\omega} \propto 1/z^2\,.
\ee
The lift~(\ref{weyl}, \ref{ansatz}) gives (after appropriate rescalings)
\be
ds^2 = z^{2 \lambda \over 4 + \lambda } \left( -{ dt^2 \over z^{  16 \over 4 + \lambda }} + {dz^2 + dy_1^2 + dy_2^2 \over z^2} \right) \,,
\ee
conformal to a $z={ 8 \over 4 + \lambda}$ Lifshitz spacetime. {The standard uplift to regular four dimensional gravity sets $\lambda = 2$ giving a dynamical exponent of $z = {4 \over 3}$.} The toy model has the same qualitative features as the magnetic brane model.  The prepotential has a single maximum $\OL W$.  For $c_2 = \OL W$, the geometry flows from conformal Lifshitz to $AdS_2$, for $c_2 > \OL W $ there is a naked singularity, and for $c_2 < \OL W $ the solution is a black hole.

The toy model arises from no known reduction, so the conformal Lifshitz geometry is a fiction.  The important point is that for both the magnetic brane and toy models, the asymptotic behavior regulates the backreaction so that the latter can be studied in a controlled way.  As we have seen, the toy model model has the same qualitative features as the magnetic brane model, in particular a flow to an IR $AdS_2$ geometry.  As its dynamical equations are simpler we will focus our attention on it.

\sect{The $\lambda = 0$, $U(\Phi) = C-A \Phi^2 $ model}

We consider in this section the $\lambda = 0$ model with the dilaton potential given by $U(\Phi) = C-A \Phi^2$, with $A, C$ positive.  By rescaling fields and coordinates we set the constants to the convenient values $A= C=2$.
We take the matter action be independent of the dilaton, $\Omega = 1$. The equations of motions are
\begin{align}
4 \partial_+ \partial_- f &= 0  \,,\label{eom:scalar}\\
2 \partial_+ \partial_- \Phi^2+ e^{2 \omega}  \left( \Phi^2 - 1\right) &= 0 \,,\label{eom:dilaton}\\
4 \partial_+ \partial_- \omega +  {e^{2 \omega} } &= 0 \,,\label{eom:metric}\\
-e^{2 \omega} \partial_+ \left( e^{-2 \omega} \partial_+ \Phi^2 \right) &= {1 \over 2} \partial_+ f \partial_+ f \,,\label{eom:constraint+}\\
-e^{2 \omega} \partial_- \left( e^{-2 \omega} \partial_- \Phi^2 \right)& = {1 \over 2} \partial_- f \partial_- f \,.\label{eom:constraint-}
\end{align}

\subsection{Vacuum solutions}

Eq.~(\ref{eom:metric}) for $\omega$ decouples from $\Phi$ and $f$, and describes a spacetime of constant negative curvature.  In Poincar\'e coordinates the solution is
\be
e^{2\omega} = \frac{1}{z^2}  = \frac{4}{(x^+ - x^-)^2} \,. \label{met1}
\ee 
The general vacuum solution for $\Phi^2$ is then given by integrating the constraints~(\ref{eom:constraint+},\ref{eom:constraint-}) and then imposing the equation of motion~(\ref{eom:dilaton}):
\be
\Phi^2 = 1 + \frac{a + b(x^+ + x^-) + cx^+ x^-}{x^+ - x^-} \,. \label{phi1}
\ee
If $ac - b^2 \neq 0$, we can bring this by an $SL(2,R)$ transformation to the form
\be
\Phi^2 = 1 + \frac{1  - \mu x^+ x^-}{x^+ - x^-} \,, \label{phi2}
\ee
where the coordinates are now dimensionless.  
More generally we will sometimes consider
\be
\Phi^2 = 1 + \frac{a  - \mu x^+ x^-}{x^+ - x^-} \, ,\label{phi3}
\ee
which allows us to continue to the pure $AdS_2$ case $a=0$; $z=a$ is the transition between conformal Lifshitz and $AdS_2$ behavior.

\begin{figure}[!t]
    \begin{center}  
     \includegraphics[width=8cm]{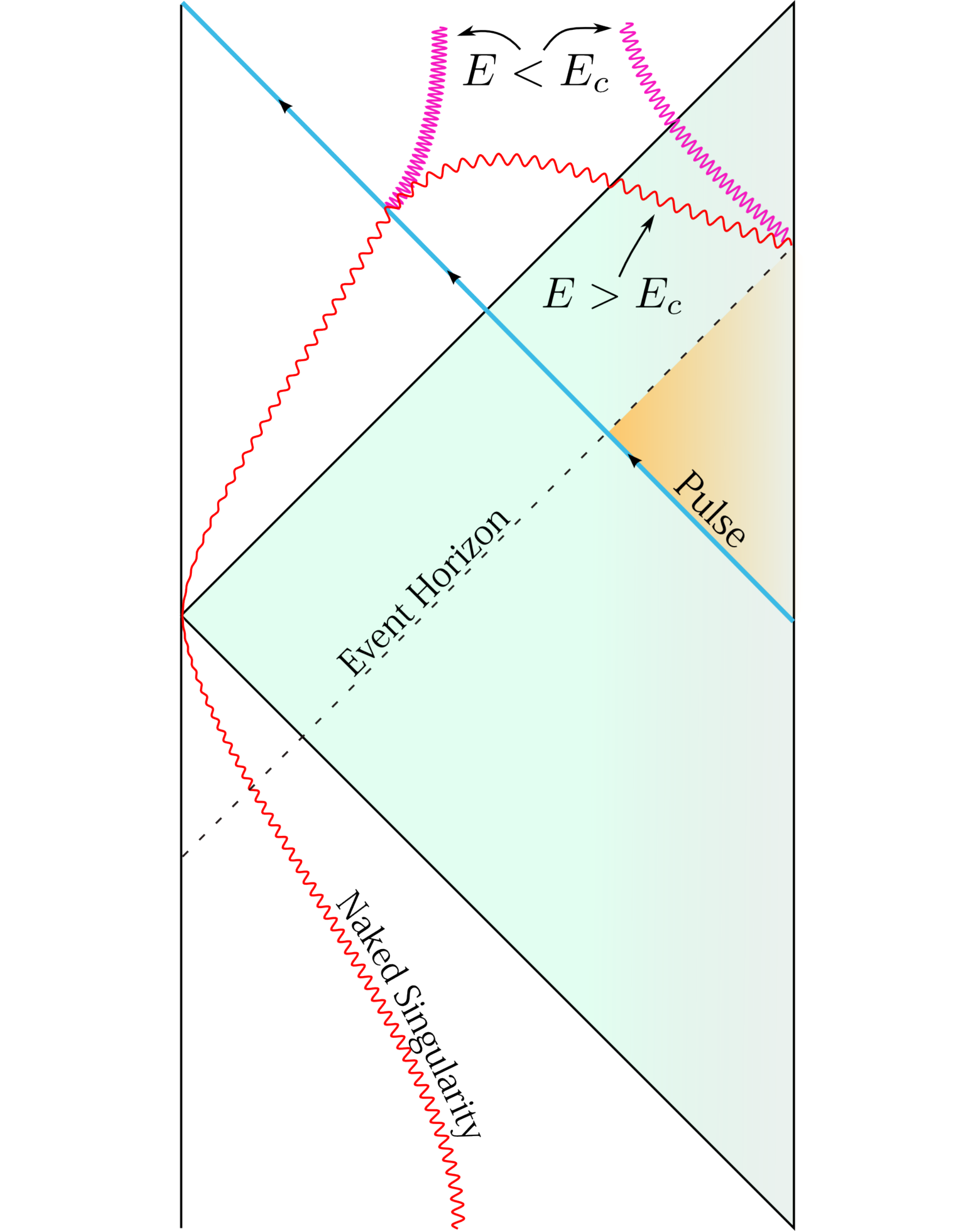}  
    \end{center}
 \caption{The global spacetime for an ingoing matter pulse ($z$ increases to the left).  Before the pulse the solution is the vacuum geometry~(\ref{glovac}), with a singularity outside the Poincar\'e patch.  After the pulse there is the black hole (\ref{phi2}).   For pulses of small amplitude, the singularity remains timelike and a second timelike singularity emerges from the boundary.  For pulses of large amplitude, the singularity is spacelike. The orange shaded portion is the exterior of the blackhole. \label{globalpic}}
  \hfill
\end{figure}

The most general vacuum solution is a conformal transformation of~(\ref{phi2}), 
\be
e^{2\omega} = \frac{4\partial_+ w^+(x^+) \partial_- w^-(x^-)}{\left( w^+(x^+) - w^-(x^-) \right)^2} \,,\quad
\Phi^2 = 1 + \frac{1  - \mu w^+(x^+) w^-( x^-)}{w^+(x^+) - w^-(x^-)} \,, \label{genvac}
\ee
for general monotonic $w^+(x^+), w^-(x^-)$.
For $\mu = 0$ the solution (\ref{met1},\ref{phi2}) interpolates from conformal Lifshitz at $z = 0$ to $AdS_2$ at large $z$, as described in \S2.
This can be converted to global coordinates via $w^\pm(x^\pm) = \tan x^{\pm}$, giving
\be
e^{2\omega} = \frac{4}{\sin^2(x^+ - x^-)} \,,\quad
\Phi^2 = 1 + \frac{\cos x^+ \cos x^-}{\sin(x^+ - x^-)} \,.
\label{glovac}
\ee
The extended geometry is shown in Fig.~\ref{globalpic}.  The $AdS_2$ behavior holds only in the neighborhood of the Poincar\'e horizons.  The metric represents global $AdS_2$, but the dilaton is nonstatic and it goes to zero in the complementary Poincar\'e patch:  in this patch the dilaton is simply given by $z \to - \tilde z$, i.e. $\Phi = 1 - 1/2 \tilde z$.   In the four dimensional lift this zero is a curvature singularity. 

For positive $\mu$ the solution~(\ref{phi2}) is a black hole.  Its mass is $\mu/8\pi G$, as will be seen in Eq.~(\ref{pulsol}) from the response to a matter pulse, and in Eq.~(\ref{energy}) from a calculation of the ADM mass.  The solution can be converted to the static form~(\ref{statmets}) by the conformal transformation $w^\pm(x^\pm) = \mu^{-1/2}\tanh \mu^{1/2}x^{\pm}$, giving
\be
e^{2\omega} = \frac{4\mu}{\sinh^2\mu^{1/2}(x^+ - x^-)} \,,\quad
\Phi^2 = 1+ \mu^{1/2}\coth\mu^{1/2}(x^+ - x^-) \label{statbh}
\ee
The coordinates~(\ref{phi2}) cover the whole black hole geometry, while the coordinates~(\ref{statbh}) cover only the exterior.  In~(\ref{phi2}), the horizon is the null line $x^+ = - \mu^{-1/2}$, while 
the singularity is at
\be
(x^+ + 1/\mu)(x^- - 1/\mu) = (\mu - 1)/\mu^2 \,.
\ee
For $\mu < 1$, the singularity remains timelike and a second timelike singularity appears near $z = 0$.  For $\mu > 1$ the singularity is spacelike.  This is shown in Fig.~\ref{globalpic}.  The singularity is naked in the global vacuum solution, but for any positive $\mu$ it is behind the black hole horizon.  In Schwarzschild coordinates, the black hole solution is
\be
ds^2 = - 4 (\rho^2 - \mu) dt^2 + \frac{ d\rho^2}{\rho^2 - \mu} \,, \quad \Phi^2 = 1 + \rho \,.
\ee
The Hawking temperature is
\begin{align}
T = -{1 \over 4 \pi} \partial_\rho \sqrt{-\frac{ g_{t t} }{g_{\rho\rho}}} \bigg|_{\rho \rightarrow \sqrt{\mu}} = {\sqrt{\mu} \over  \pi} \,.  \label{temp}
\end{align}

\subsection{Solutions with matter}

One of the main attractions of this model is that the equations of motion are linear in $\Phi^2$, so it remains solvable with matter.  
In coordinates with metric~(\ref{met1}), $e^{2\omega} = {4}/{(x^+ - x^-)^2}$, the constraints~(\ref{eom:constraint+},\ref{eom:constraint-}) can be written as
\be
\partial_\pm \partial_\pm M(x^+, x^-) = - (x^+ - x^-) 8\pi G T_{\pm\pm}(x^\pm)\,,
\ee
where $\Phi^2 = M/(x^+ - x^-) $ and $T_{\pm\pm} = \partial_\pm f \partial_\pm f/16\pi G$.
The general solution, integrating from some reference point $u$, is then
\begin{align}
M &= M_0 - I^+ + I^- \,,
\end{align}
where
\begin{align}
I^\pm(x^+, x^-) &= 8\pi G \int_{u^\pm}^{x^\pm} dx'^\pm\,
(x'^\pm - x^\mp)(x'^\pm - x^\pm) T_{\pm\pm}(x'^\pm) \,,
\end{align} 
and $M_0$ is any sourceless solution,
\be
M_0 = a + bx^+ + cx^- + dx^+ x^-\,, \quad b - c = 2\,.
\ee


For example, start with the Poincar\'e vacuum in the form~(\ref{phi2}) and throw in a pulse of energy $E$,
\begin{equation}
T_{--}  = E  \,\delta(x^-) \,,
\end{equation}
which describes a shockwave traveling on the null curve $x^- = 0$, emanating from the boundary.   The solution is then
\be
\Phi^2 = 1 + \frac{a - 8\pi G E \theta(x^-) x^+ x^-}{x^+ - x^-} \,, \label{pulsol}
\ee
describing formation of a black hole with $\mu = 8\pi GE$, as in Fig.~\ref{globalpic}. As above, after the pulse there are timelike singularities for $E < E_c = 1/8 \pi G a$, and a spacelike singularity for $E > E_c$.

\subsection{Backreaction in $AdS_2$}

For the solution~(\ref{phi3}), the black hole singularity reaches the boundary at $t = \pm (a/\mu)^{1/2}$.  The boundary conditions in the exterior region $-(a/\mu)^{1/2} < t < (a/\mu)^{1/2}$ remain conformal Lifshitz.  However, for $a=0$, where the vacuum solution is pure $AdS_2$, no region of the boundary remains: the past and future singularities meet at the boundary point $t=0$.  Similarly, when any pulse is thrown into the $a=0$ $AdS_2$ geometry, a singularity forms instantly on the boundary and no part of the $AdS_2$ boundary survives.  Thus $AdS_2$, with a finite volume transverse space admits no finite energy excitations~\cite{Maldacena:1998uz}.

\sect{Scalar field holography}

We will use this model to investigate the effect of backreaction on boundary scalar correlation functions in both the vacuum and black hole backgrounds.  Previous investigations of $AdS_2$ holography~\cite{ads2holo} have studied either propagating fields or the gravitational sector separately, whereas we are interested in the coupling between the two.

We continue to Euclidean coordinates; the fields remain real. Working in this regime also avoids the singularity that appeared in the global picture, which is completely excised in the Euclidean geometry. We take the continuation
\begin{align}
&x^+ = t + z \rightarrow -i \tau + z \equiv x \,, \nonumber\\
&x^- = t - z \rightarrow -i \tau - z \equiv - \bar{x} \,.
\end{align}

The only propagating field is $f$, whose normalizable and nonnormalizable solutions scale respectively as $z^1$ and $z^0$.
We will introduce a boundary source for the nonnormalizable mode,
\begin{equation}
\lim_{z \to 0} f(z,\tau) = j(\tau)\,, \label{boundc}
\end{equation}
to produce a generating functional for correlators of the corresponding boundary operator.  The metric is again unaffected by the source,
\begin{align}
ds^2 = \frac{4 dx d\bar x}{(x + \bar x)^2} \,.
\label{euclidean:metric}
\end{align}
For the dilaton we take the asymptotic behavior
\be
\Phi^2(x,\bar x) \stackrel{{\rm large}\ x}{\to} \Phi_0^2(x,\bar x) = 1 + \frac{a + \mu x\bar x}{x+\bar x} \,.
\label{euclidean:dilaton}
\ee

The generating functional can be obtained by plugging the asymptotic solutions into the action boundary term. Using the equations of motion, the on-shell action (\ref{fullact}) regulated at $z = \epsilon$ is
\begin{align}
S_{\rm reg} &= -{1 \over 8 \pi G} \int dt dz \,  e^{2 \omega} + {1 \over 8 \pi} \int_\epsilon dt \left(  {\Phi^2 \over \epsilon} + {1 \over 4} f \partial_z f  \right) \\
&= {1 \over 8 \pi G} \int_\epsilon dt \left( - {1 \over  \epsilon}  + {\Phi^2 \over \epsilon} + {1 \over 4} f \partial_z f  \right)  \,.
\end{align}
The divergences from removing the regulator are canceled by the following counterterms
\begin{align}
S_{\rm ct} &= {1 \over 8 \pi G}  \int_\epsilon dt \ \left( - e^{\omega} \Phi^2  +  e^{\omega}\right) \\
&= {1 \over 8 \pi G}  \int_\epsilon dt \ \left( - {\Phi^2 \over \epsilon}  + {1 \over \epsilon} \right)  \,.
\end{align}
Thus the renormalized action is just the boundary term
\begin{align}
S_{\rm ren} &= {1 \over 32 \pi G} \int dt \, f \partial_z f  \,. \label{action:ren}
\end{align}
To evaluate this, the bulk-to-boundary propagator with boundary condition~(\ref{boundc}) gives
\be
f(z, \tau) = \frac{1}{2\pi} \int_{-\infty}^\infty d\tau' \left(\frac{1}{x + i\tau'} + \frac{1}{\bar x - i\tau'} \right) j(\tau') \,.
\label{fbulk}
\ee
where $x = z - i\tau$.  Then 
\begin{align}
\lim_{z \to 0} f(z, \tau)  &= j(\tau) \,, \nonumber \\
\lim_{z \to 0} \partial_z f(z, \tau)  &= \frac{1}{\pi} \int_{-\infty}^\infty d\tau' \frac{ P}{\tau - \tau'}\, \partial_{\tau' } j(\tau') \,,\label{fdf}
\end{align}
with $P$ the principal part.  The action~(\ref{action:ren}) is then
\be
S_{\rm ren} = - {1 \over 32 \pi G} \int_{-\infty}^\infty  d\tau\, d\tau' \, \frac{ P}{(\tau - \tau')^2} \, j(\tau) j(\tau')  \,.
\ee

This result is superficially plausible.  In the IR $AdS_2$ region it corresponds to a conformal operator of dimension~1, and this behavior continues into the UV regime because $f$ does not couple to the transverse metric $\Phi^2$.  However, it cannot be the whole story.  First, it is insensitive to the black hole mass $\mu$, which should break the conformal invariance.  Second, the result is gaussian, there are no interactions, but we have argued in the previous section that backreaction has large effects.

The subtlety is that we must correctly relate the time $\tau$ to the time in the dual field theory.  In the coordinates we have been using, the asymptotic behavior of  $\Phi$ is deformed, and we must transform back to coordinates with standard asymptotics.  This will introduce a dependence on the black hole mass, and also a highly nonlinear dependence on the scalar fields. 

The Euclidean solution for $\Phi^2$ is
\begin{align}
\Phi^2 &= \Phi^2_0 - {I + I^* \over x + \bar{x}} \,,
\end{align}
where
\begin{align}
I(x, \bar x) &= 8\pi G \int_u^x dx'\,
(x' + \bar x)(x' - x) T(x') 
\end{align} 
and $\Phi_0$ is any sourceless solution.  For $j$ of compact support, the bulk $f$ (\ref{fbulk}) falls as $1/|x|^2$, and so the integral $I$ converges as $|u| \to \infty$.  It is convenient to set $u = i \infty$ (i.e. $\tau_u = -\infty$), and let 
\be
\Phi_0^2 = 1 + \frac{a + \mu x\bar x}{x+\bar x} \,.
\ee
Then the asymptotic behavior of $\Phi^2$ is
\begin{align}
M(\tau) \equiv \,\lim_{z \to 0}\, (x + \bar x) \Phi^2 &= a + \mu \tau^2 + 16 \pi G \int_{-\infty}^\tau d\tau'\, (\tau' - \tau)^2 \, {\rm Im}\, T(-i\tau') 
\nonumber\\
&= a + \mu \tau^2 + 4  \int_{-\infty}^\infty d\tau' d\tau''\, H(\tau, \tau',\tau'')
\partial_{\tau'}j(\tau') \partial_{\tau''}j(\tau'')  \,,
\end{align}
where
\begin{align}
H(\tau, \tau',\tau'') = \frac{(\tau'-\tau)^2  \theta( \tau-\tau') - (\tau''-\tau)^2  \theta(\tau-\tau'')   }{ \tau' - \tau''} \,.
\end{align}

In order to bring the dilaton back to fixed asymptotic behavior, while keeping the asymptotic form of the metric, we need a new coordinate $\tilde x(x) = \tilde z - i\tilde \tau$ such that 
\be
\frac{M(\tau)}{x + \bar x} \approx \frac{a}{\tilde x + \bar{\tilde x}} \,.
\ee
This implies the differential equation
\be
\frac{\partial \tilde \tau}{\partial \tau} = \frac{a}{M(\tau)}\,. \label{timediff}
\ee
It is $\tilde\tau$ that is to be identified with the time in the boundary field theory.

Let us illustrate this for the situation that the backreaction can be neglected compared to the effect of the black hole mass, so $M = a + \mu \tau^2$.  Then $\tau = (a/\mu)^{1/2} \tan[ (\mu/a)^{1/2}\tilde\tau]$.  The bulk field $f$ is a scalar, so its boundary limit transforms
\be
\tilde\jmath(\tilde\tau) = j(\tau) \,.
\ee
The renormalized action becomes 
\be
S_{\rm ren} = - {\mu \over 32 \pi G a} \int_{-\infty}^\infty  d\tilde\tau\, d\tilde\tau' \, \frac{ P}{\sin^2 [(\mu/a)^{1/2}(\tilde\tau - \tilde\tau')]} \, \tilde\jmath(\tilde\tau) \tilde\jmath(\tilde\tau')  \,. \label{bhgen}
\ee
Then $e^{-S_{\rm ren}}$ generates the correlators as functions of the field theory time $\tilde\tau$.
These exhibit the expected nonconformality and thermal periodicity.

Now let us consider the effect of backreaction, letting $\mu = 0$ for simplicity.  The differential equation~(\ref{timediff}) can be integrated to give $\tilde\tau$ as a function of $\tau$, but this can be inverted only implicitly.  We therefore expand in $j$.  Thus
\begin{align}
\tilde\tau &= \tau -  \frac{1}{2\pi a}\int_{-\infty}^\tau d\tau'   \int_{-\infty}^\infty d\tau_1 d\tau_2 \, H(\tau', \tau_1,\tau_2)
\partial_1 j(\tau_1) \partial_2 j(\tau'_2) + O(j^4) \nonumber\\
&= \tau -  \frac{1}{6\pi a}  \int_{-\infty}^\infty d\tau_1 d\tau_2 \, \frac{(\tau-\tau_1)^3  \theta( \tau-\tau_1) - (\tau-\tau_2)^3  \theta(\tau-\tau_2)   }{ \tau_1 - \tau_2}
\partial_1 j(\tau_1) \partial_2 j(\tau_2) + O(j^4) \,.
\end{align}
To this same order,
\begin{align}
\tau &= \tilde\tau +  \frac{1}{6\pi a}  \int_{-\infty}^\infty d\tilde\tau_1 d\tilde\tau_2 \, \frac{(\tilde\tau-\tilde\tau_1)^3  \theta( \tilde\tau-\tilde\tau_1) - (\tilde\tau-\tilde\tau_2)^3  \theta(\tilde\tau-\tilde\tau_2)   }{ \tilde\tau_1 - \tilde\tau_2}
\partial_{\tilde 1} \tilde\jmath(\tilde\tau_1) \partial_{\tilde 2} \tilde\jmath(\tilde\tau_2) + O(j^4) \nonumber\\
&\equiv \tilde\tau + \gamma(\tau) + O(j^4) \,.
\end{align}
Then
\begin{align}
S_{\rm ren} &= - {1 \over 32 \pi G} \int_{-\infty}^\infty  d\tilde\tau\, d\tilde\tau' \, \frac{ P}{(\tilde\tau - \tilde\tau')^2}
\left( 1 +  \partial_\tau \gamma(\tilde\tau) +  \partial_{\tau'}\gamma(\tilde\tau') - 2\frac{\gamma(\tilde\tau) -  \gamma(\tilde\tau')}{\tilde\tau-\tilde\tau'} \right) \tilde\jmath(\tilde\tau) \tilde\jmath(\tilde\tau') + O(j^6) \nonumber\\
&\equiv - \sum_{n=1}^\infty \frac{1}{(2n)!} \int_{-\infty}^\infty  d^{2n}\tilde\tau \, G(\tilde\tau_1,\ldots,\tilde\tau_{2n}) \tilde\jmath(\tilde\tau_1) \ldots
\tilde\jmath(\tilde\tau_{2n}) \,.
\end{align}
Simplifying, we have 
\begin{align}
G(\tilde\tau_1,\tilde\tau_{2}) &= {1 \over 16 \pi G} \frac{ 1}{(\tilde\tau_1 - \tilde\tau_2)^2} \,, \nonumber\\
G(\tilde\tau_1,\tilde\tau_{2},\tilde\tau_3,\tilde\tau_{4}) &= - {1\over 24\pi^2 G a} \frac{\theta(\tilde\tau_{13})}{\tilde\tau_{12}^3\tilde\tau_{34}^3}\left(\tilde\tau_{13}^3 - 3\tilde\tau_{13}^2 \tilde\tau_{23} - 3 \tilde\tau_{23}\tilde\tau_{13}\tilde\tau_{34}\right) + {\rm 23\ permutations}\,,
\end{align}
where $\tilde\tau_{ij} = \tilde\tau_i - \tilde\tau_j$.

The expression for the four-point function can perhaps be simplified, but in any case the result is instructive.  If the theory were scale-invariant, the connected four-point function would have the same $1/\tilde\tau^4$ scaling as the disconnected one.  Instead it scales as $1/\tilde\tau^{3}$.  Thus we conclude that {\it the backreaction is a relevant interaction and explicitly breaks the scale and conformal invariance of the theory}.  Indeed, the importance of backreaction at low energy was already reached in the early work~\cite{Preskill:1991tb}.
  The connected and disconnected pieces are comparable when $\tilde{\tau} \sim a / G \propto a  V$, determining the scale where the conformal behavior breaks down {to be $E_{breaking} \sim 1/ a V$. In the present case where the dual is a field theory in finite volume, a breaking scale which decreases with volume is precisely what one expects. 
The breaking has the same scaling as a perturbation of dimension zero, but there is no candidate operator of this dimension, so evidently it cannot be interpreted, or canceled, in this way.

From another point of view, if we take $a \to 0$, this has the effect of taking the UV-$AdS_2$ transition to infinite energy, producing a pure $AdS_2$ theory.  We see that the four-point function diverges in this limit, so there is no sensible $AdS_2$ CFT.  These results are plausible, given that the backreaction allows no finite energy states.  They have been derived only in the solvable model, but we expect that they are a general property of $AdS_2 \times$compact solutions. 
If we go beyond the classical limit to consider bulk loop corrections to the two-point functions, the relevant interactions will produce large corrections to the IR behavior; it would be interesting to study these.

\section{The black hole and matter in equilibrium}
\setcounter{equation}{0}

The main motivation for the CGHS model was the study of black hole evaporation through quantum production of $f$ quanta~\cite{Callan:1992rs}.  We do not expect the models considered here to differ substantively in this regard.  Our interest here is to study the contribution of the matter fields to thermodynamic properties, in connection with the effect of backreaction. 

\subsection{Backreaction of matter}

For $N$ scalar fields, the conformal anomaly is $T^\mu\!_\mu =  N R/24\pi$, or
\be
T_{+-} = -\frac{N}{12\pi} \partial_+ \partial_-  \omega\,. \label{tpm}
\ee
This expression is not invariant under the Weyl transformation~(\ref{weyl}).  The ambiguity corresponds to the possible addition of a term $R \ln \Phi$ to the Lagrangian.  We define the model so that the form~(\ref{tpm}) holds in the frame in which $\lambda = 0$, which simplifies the equations of motion.  Again, in the $AdS_2$ region of interest the dilaton is constant and the different choices become equivalent.

Conservation of $T_{ab}$ implies that
\be
T_{\pm\pm} = \frac{N}{12\pi} (\partial_\pm^2  \omega - \partial_\pm \omega\partial_\pm \omega) + \tau_{\pm\pm}(x^\pm) \,.
\ee
The equations of motion become
\begin{align}
2 \partial_+ \partial_- \Phi^2+ e^{2 \omega}  \left( \Phi^2 - 1\right) &= 16\pi G T_{+-} \,,\label{xeom:dilaton}\\
4 \partial_+ \partial_- \omega +  {e^{2 \omega} } &= 0 \,,\label{xeom:metric}\\
-e^{2 \omega} \partial_+ \left( e^{-2 \omega} \partial_+ \Phi^2 \right) &=  8\pi G T_{++} \,,\label{xeom:constraint+}\\
-e^{2 \omega} \partial_- \left( e^{-2 \omega} \partial_- \Phi^2 \right)& = 8\pi G T_{--}  \,.\label{xeom:constraint-}
\end{align}
The equation for the metric is unaffected, so we take again the static black hole solution $e^{2\omega} = {4\mu}/{\sinh^2 2\mu^{1/2}z}$.
For a static solution the remaining equations become
\begin{align}
(\Phi^2)'' &= \frac{8 \mu}{\sinh^2 2\mu^{1/2}z} (\Phi^2 - 1 - GN/3) \\
(\sinh^2 2\mu^{1/2}z (\Phi^2)')' &= 32\pi G (\tau - N/12\pi) \sinh^2 2\mu^{1/2}z  \label{yeom:constraint}\,,
\end{align}
where $\tau_{++} = \tau_{--} = \tau$.  If $\tau - N/12\pi \neq 0$, the constraint~(\ref{yeom:constraint}) implies that $\Phi^2$ diverges $\propto z$ as we approach the horizon $z \to \infty$.  Thus $\tau = N/12\pi$ and $T_{\pm\pm}$ must vanish identically, a curious result.  The solution is then
\be
\Phi^2 = 1 + \frac{GN}{3} + \mu^{1/2}  \coth 2\mu^{1/2}z \,.
\ee
The large-$N$ quantum effect is just a constant shift of $\Phi^2$.

The inclusion of the large-$N$ quantum effects actually has no effect on the holographic correlators of \S4.  The metric equation~(\ref{xeom:metric}) is unaffected, and the shift of $\Phi^2$ is only in the subleading term and so does not alter the time reparameterization.  However, the thermodynamic properties depend on $N$, as we now show.

\subsection{The renormalized stress tensor}

In order to obtain the energy-temperature relation for the black hole, we will compute the boundary stress tensor.  To convert this to a classical problem, we replace the $f$ fields with an equivalent classical system.  The coupling of a conformal system to gravity is determined only by the central charge, and so we replace the $f$ fields with a single $\chi$ with action
\begin{align}
S_\chi =  -{N \over 24 \pi} \int d^2x \sqrt{-g} \left(  \partial_\mu \chi \partial^\mu \chi +  \chi R\right)
- {N \over12 \pi} \int dt \sqrt{-\gamma} \chi K\,, \label{Qcorr}
\end{align}
giving central charge $N$ (we work in the large-$N$ approximation, ignoring loops of $\chi$).  The full renormalized action is
\begin{align}
S_R &= S_G + S_\chi + S_{\rm ct} \,,\nonumber\\
S_G &= {1 \over 16 \pi G} \int d^2x \sqrt{-g}\left( \Phi^2 R -  U_{0}(\Phi) \right)  + {1 \over 8 \pi G} \int dt \sqrt{-\gamma} \Phi^2  K \,.
\end{align}
The equations of motion in conformal gauge are again (\ref{xeom:dilaton}-\ref{xeom:constraint-}), where now
\begin{align}
T_{+-} &= \frac{N}{12\pi} \partial_+ \partial_- \chi  \,, \nonumber\\
T_{\pm\pm} &= \frac{N}{12\pi} \left( - \partial_\pm^2 \chi + \partial_\pm \chi \partial_\pm\chi + 2 \partial_\pm \omega \partial_\pm \chi \right)   \,,\nonumber\\
\partial_+ \partial_- (\chi + \omega) &= 0 \,.
\end{align}
The solution
\begin{align}
e^{2\omega} &= {4\mu}/{\sinh^2 2\mu^{1/2}z}\,, \quad \chi = -\omega -2\mu^{1/2} z \,, \nonumber\\
\Phi^2 &= 1 + \frac{GN}{3} +  \mu^{1/2}  \coth 2\mu^{1/2}z \,. \label{chisol}
\end{align}
reproduces that in \S5.1, with $\chi$ going to a constant on the horizon.

We now obtain the counterterm action.  In conformal gauge,
\be
S_G + S_\chi = \int dz\,dt\, \left\{ \frac{1}{8\pi G}\left( -4 \partial_{(+}\Phi^2 \partial_{-)}\omega + (\Phi^2 - 1) e^{2\omega} \right)
+ \frac{N}{6\pi} \left( \partial_+ \chi \partial_- \chi + 2 \partial_{(+}\chi \partial_{-)}\omega \right) \right\}\,.
\ee
Inserting the asymptotic expansion of the solution (\ref{chisol}),
\begin{align}
e^{2\omega} &= \frac{1}{z^2} - \frac{4\mu}{3} + O(z^2) \,, \nonumber\\
\Phi^2 &= \frac{1}{2z} + 1 + \frac{GN}{3} + \frac{2\mu z}{3} + O(z^3) \,, \nonumber\\
\chi &= \ln z - 2\mu^{1/2} z + \frac{2\mu z^2}{3} + O(z^4) \,,  \label{asymp}
\end{align}
the divergent part of the action is 
\be
S_G + S_\chi = \frac{1}{16\pi G \epsilon^2} + \frac{N}{12\pi \epsilon} + {\rm finite} \,.
\ee
This is canceled by the local counterterms
\be
S_{\rm ct} = \int dt\,  \sqrt{-\gamma}\left\{\frac{1}{8\pi G} (1 - \Phi^2) - \frac{N}{24\pi} \right\}\,.
\ee

Next we compute the boundary stress tensor following the approach of Ref.~\cite{Balasubramanian:1999re}, varying with respect to the boundary metric. The prescription is given by
\begin{align}
\langle \hat T_{tt} \rangle = -{2 \over \sqrt{-\hat\gamma}} {\delta S_R \over \delta \hat\gamma^{tt}} = \lim_{\epsilon \rightarrow 0}{-2\epsilon \over \sqrt{-\gamma(\epsilon)}} {\delta S_R(\epsilon) \over \delta \gamma^{tt}(\epsilon)} \,. \label{ttt}
\end{align}
with $S_R$ is the renormalized action.  Note that hats refer to the dual field theory, so that $\hat\gamma_{tt} = \lim_{\epsilon \to 0} \epsilon^2 \gamma_{tt}(\epsilon)$ is the metric of the boundary theory. The Hamilton-Jacobi formalism gives the functional derivative as
\be
{\delta S_R(\epsilon) \over \delta \gamma^{tt}(\epsilon)} = -\pi_{tt}(\epsilon) + {\delta S_{\rm ct}(\epsilon) \over \delta \gamma^{tt}(\epsilon)}  \,,  \label{dsr}
\ee
where
\be
\pi_{tt} = \frac{\partial(L_G + L_\chi)}{\partial (\partial_z g^{tt})}   \,.
\ee
The relevant terms in the action are 
\begin{align}
S_G + S_\chi &\approx  \int d^2x \sqrt{-g} X R
+ \int dt \sqrt{-\gamma} X K\,,  \nonumber\\
&= \int d^2x \sqrt{-g} \partial^\mu X (g_{\mu\rho} \partial_\sigma - g_{\sigma\rho} \partial_\mu) g^{\sigma\rho}  \,,
\end{align}
where $X = \Phi^2/16\pi G - N\chi/24\pi$.  Thus we read off 
\be
\pi_{tt} = -  \sqrt{-g} g_{tt} \partial^z X = e^{2\omega} \partial_z X \,. \label{pitt}
\ee
Combining Eqs.~(\ref{ttt},\ref{dsr},\ref{pitt}), we have
\be
\langle \hat T_{tt} \rangle =  2 \epsilon e^\omega \partial_z X - \epsilon e^{2\omega} \left( \frac{1-\Phi^2}{8\pi G} - \frac{N}{24\pi} \right) \,.
\ee
Inserting the asymptotic expansions~(\ref{asymp}) then gives
\be
\langle \hat T_{tt} \rangle = \frac{\mu }{8\pi G} + \frac{N \mu^{1/2}}{6\pi}  \,.  \label{energy}
\ee

\subsection{Thermodynamic quantities}

The metric is unaffected by the coupling to matter, so the temperature~(\ref{temp}) is as before, $T = \mu^{1/2} / \pi$.
The equation of state is then 
\be
E = \frac{\pi}{8 G}  T^2 + \frac{N}{6}T \,. \label{energ}
\ee
From this we can compute the entropy, using $dS = dE/T$:
\begin{align}
S = \frac{\pi}{4 G}   T + { N \over 6} \ln{T} + c \,.  \label{sholo}
\end{align}
It is interesting to compare this with the Bekenstein-Hawking entropy.\footnote{The same $(N/6) \ln T$ term was previously found as an entanglement entropy in Ref.~\cite{Spradlin:1999bn}.  We thank N. Iqbal for bringing this to our attention.}

 {We can read off the effective gravitational constant from the coefficient of the Ricci scalar in the actions~(\ref{fullact}) and (\ref{Qcorr}) as $1/G_{eff} = {\Phi^2 / G} - {2 N \chi / 3}$, and with the horizon a point of area 1, we have
\begin{align}
S_{\rm BH} = {\Phi^2 \over 4 G} - {N \chi\over 6} \biggr|_{z \to \infty} &= {\sqrt{\mu} \over 4 G} + {1 \over 4 G} + {N\over6}\ln{4 \sqrt{\mu}} + {N \over 12} \nonumber \\
&= {\pi T \over 4 G} + {1 \over 4 G} + {N\over6}\ln{T} + {N \over 6}\ln{4 \pi} +  {N \over 12}  \label{sbh}
\end{align}We get agreement between the holographic~(\ref{sholo}) and Bekenstein-Hawking~(\ref{sbh}) entropies, with 
\begin{align}
c =  {1 \over 4 G}+ {N \over 6}\ln{4 \pi} +  {N \over 12} \,.
\end{align}}
In interpreting the holographic entropy, we should note that the closely related CGHS model describes remnants~\cite{Banks:1992ba,Giddings:1992ff,Almheiri:2013wka} .  Since an arbitrarily large black hole can decay down to a Planck mass remnant, the number of states at low energy is unbounded above, and it is unrelated to the thermodynamic Bekenstein-Hawking entropy.  Presumably the same is true for the model that we are considering, taken on its own terms as a bulk quantum field theory.  However, we are merely using this model as an approximation to the behavior of precise gauge/gravity duals such as those of Ref.~\cite{Almheiri:2011cb}.  For these, we can be fairly confident that the dual field theory dynamics do not allow an unbounded number of states at finite energy, and expect rather that the thermodynamic and statistical entropies agree.  We will therefore interpret the holographic entropy as representing the true density of states.

Our work was motivated in part by a puzzle regarding the density of states in $AdS_2/CFT_1$ duals~\cite{Jensen:2011su,Iqbal:2011in}.  Our result exhibits a related puzzle: the entropy~(\ref{sholo}) becomes negative at sufficiently low temperature, due to the log term.  Our discussion of backreaction suggests that we cut this off where the theory becomes strongly coupled, at $T \sim G$.  Effectively we are using backreaction to provide the cutoff introduced by hand in Ref.~\cite{Iqbal:2011in}.

Thus replace $\ln T$ with $\ln (T+G)/G$, so that the log goes smoothly to zero as $T \to 0$ (for $T \gg G$ this means that 
 $c \sim - (N/6) \ln G $).  However, this is not fully satisfactory as the entropy is no longer extensive: with $G\propto 1/V$ and $N \propto V$, there is a $V \ln V$ term.  We might cancel this by an additional term $- (N/3) \ln N$ but this seems rather ad hoc.  Or it may be that our reduced model, which retains only the zero mode of the gravitational field, simply fails to incorporate extensivity.
 
{One ingredient missing in this analysis is the effect of bulk loops on the backreaction scale. This includes processes with $N$ scalars running in loops that have the potential of significantly ramping up the conformal symmetry breaking scale. Preliminary results \cite{loops} indicate that this is indeed the case and that the new backreaction scale is pushed up to $T \sim N G$. Using this to cut-off the $\ln T$ as $\ln(T + N G)/N G$ gets the job done and preserves extensivity. Performing the full analysis to confirm this prediction will be an interesting problem to pursue.}

Let us also reiterate the puzzle of Refs.~\cite{Jensen:2011su,Iqbal:2011in}.  In the noncompact limit, one expects the conformal symmetry to be exact.  However, in the far infrared, the only conformally invariant behavior for the entropy is $T^0$, from the well-known zero-energy degeneracy.  But if all states are at zero energy, how can there be any dynamics?  

It may be that backreaction provides the resolution here.  Namely, the sector of the theory that is probed by CFT $n$-point functions involves only finite numbers of excitations, whose backreaction is finite in the infinite volume limit.  On the other hand, states of finite energy density will have the same singular backreaction as in the compact case.  Thus there may be a sub-extensive set of finite energy states, whose entropy per unit volume vanishes as $V \to \infty$ but which realize the infinite-volume $AdS_2$ symmetry.
To investigate this would seem to require a more refined treatment of backreaction, going beyond the zero mode retained here.

\section{Discussion}

We have studied models of 1+1 dimensional gravity that flow to $AdS_2$ times a compact space in the IR.  The UV corresponds to the dimensional reduction of a higher dimensional scale invariant theory, which regulates the $AdS_2$ backreaction.  In particular, we have focused on a model in which the backreaction is solvable, as in the CGHS model.  An interesting result was  the calculation of holographic correlators, and the demonstration that the effect of backreaction is strongly relevant in the $AdS_2$ region.  

We have argued that the solvable model, although it does not result from any specific reduction, nevertheless captures universal behaviors.  For more general models, it will be likely necessary to solve numerically. 

Given the ubiquity of $AdS_2$ spacetimes and the importance of their backreaction, we hope that our model will be useful.
For applications to condensed matter systems, the transverse space is generally noncompact.  Backreaction may still be important to 
understanding the density of states, as we have discussed.  Also, in finite density states the backreaction will be as in the compact case.

If the bulk field theory can be consistently quantized, it defines a $1+1$ dimensional conformal theory holographically.  This may be counterfactual, given the difficulty of assigning boundary conditions at the $\Phi^2 = 0$ singularity.  But supposing that it can be done, it would describe a theory of remnants~\cite{Banks:1992ba,Giddings:1992ff,Almheiri:2013wka}.  This would seem to conflict with the general lore that holography excludes remnants~\cite{Almheiri:2013hfa}.  However, the latter is based on having an explicit field theory dual that is sufficiently well understood to expect that it has a finite density of states.  Here, there is no independent definition of the dual CFT.

It would be interesting to extend the present work to include gravitational loop corrections, in particular to assess the magnitude of the corrections to our results.  Also, it would be interesting to develop a more physical interpretation of the thermodynamic quantities that we have calculated.

\section*{Acknowledgments}

We thank Nabil Iqbal, Hong Liu, Don Marolf, Ben Michel, and Eric Mintun for discussions.  J.P. was supported in part through NSF grants PHY07-57035 and PHY13-16748 (summer) and PHY11-25915 (academic year).



\begin{thebibliography}{99}
\itemsep = 3pt

\bibitem{BR}
T. Levi-Civita, R.C. Acad. Lincei 26, 519 (1917); 

B. Bertotti, Phys. Rev. 116, 1331 (1959); 

I. Robertson, Bull. Acad. Polon. 7, 351 (1959);

S.~Ferrara and R.~Kallosh,
  ``Supersymmetry and attractors,''
  Phys.\ Rev.\ D {\bf 54}, 1514 (1996)
  [hep-th/9602136].

\bibitem{Chamblin:1999tk} 
  A.~Chamblin, R.~Emparan, C.~V.~Johnson and R.~C.~Myers,
  ``Charged AdS black holes and catastrophic holography,''
  Phys.\ Rev.\ D {\bf 60}, 064018 (1999)
  [hep-th/9902170].
  
\bibitem{Iqbal:2011in} 
  N.~Iqbal, H.~Liu and M.~Mezei,
  ``Semi-local quantum liquids,''
  JHEP {\bf 1204}, 086 (2012)
  [arXiv:1105.4621 [hep-th]].
  
\bibitem{Maldacena:1998uz} 
  J.~M.~Maldacena, J.~Michelson and A.~Strominger,
  ``Anti-de Sitter fragmentation,''
  JHEP {\bf 9902}, 011 (1999)
  [hep-th/9812073].
  
\bibitem{Callan:1992rs} 
  C.~G.~Callan, Jr., S.~B.~Giddings, J.~A.~Harvey and A.~Strominger,
  ``Evanescent black holes,''
  Phys.\ Rev.\ D {\bf 45}, 1005 (1992)
  [hep-th/9111056].
  
 \bibitem{Hartnoll:2007ai} 
  S.~A.~Hartnoll and P.~Kovtun,
  Phys.\ Rev.\ D {\bf 76}, 066001 (2007)
  [arXiv:0704.1160 [hep-th]].
  
  \bibitem{D'Hoker:2009mm} 
  E.~D'Hoker and P.~Kraus,
  ``Magnetic Brane Solutions in AdS,''
  JHEP {\bf 0910}, 088 (2009)
  [arXiv:0908.3875 [hep-th]].
  
\bibitem{Donos:2011pn} 
  A.~Donos, J.~P.~Gauntlett and C.~Pantelidou,
  ``Magnetic and Electric AdS Solutions in String- and M-Theory,''
  Class.\ Quant.\ Grav.\  {\bf 29}, 194006 (2012)
  [arXiv:1112.4195 [hep-th]];
  
  A.~Donos and J.~P.~Gauntlett,
  ``Supersymmetric quantum criticality supported by baryonic charges,''
  JHEP {\bf 1210}, 120 (2012)
  [arXiv:1208.1494 [hep-th]].
    
\bibitem{Almheiri:2011cb} 
  A.~Almheiri,
  ``Magnetic AdS2 x R2 at Weak and Strong Coupling,''
  arXiv:1112.4820 [hep-th].
  
\bibitem{Sen:2008vm} 
  A.~Sen,
  ``Quantum Entropy Function from AdS(2)/CFT(1) Correspondence,''
  Int.\ J.\ Mod.\ Phys.\ A {\bf 24}, 4225 (2009)
  [arXiv:0809.3304 [hep-th]];
  ``State Operator Correspondence and Entanglement in $AdS_2/CFT_1$,''
  Entropy {\bf 13}, 1305 (2011)
  [arXiv:1101.4254 [hep-th]].
  
\bibitem{deAlwis:1992hv} 
  S.~P.~de Alwis,
  ``Quantum black holes in two-dimensions,''
  Phys.\ Rev.\ D {\bf 46}, 5429 (1992)
  [hep-th/9207095];

  A.~Bilal and C.~G.~Callan, Jr.,
  ``Liouville models of black hole evaporation,''
  Nucl.\ Phys.\ B {\bf 394}, 73 (1993)
  [hep-th/9205089];
  
  S.~B.~Giddings and A.~Strominger,
  ``Quantum theories of dilaton gravity,''
  Phys.\ Rev.\ D {\bf 47}, 2454 (1993)
  [hep-th/9207034].

\bibitem{Grumiller:2002nm} 
  D.~Grumiller, W.~Kummer and D.~V.~Vassilevich,
  ``Dilaton gravity in two-dimensions,''
  Phys.\ Rept.\  {\bf 369}, 327 (2002)
  [hep-th/0204253].
  
\bibitem{Giddings:1992kn} 
  S.~B.~Giddings and A.~Strominger,
  ``Dynamics of extremal black holes,''
  Phys.\ Rev.\ D {\bf 46}, 627 (1992)
  [hep-th/9202004].

\bibitem{Kachru:2009xf} 
  S.~Kachru, A.~Karch and S.~Yaida,
  ``Holographic Lattices, Dimers, and Glasses,''
  Phys.\ Rev.\ D {\bf 81}, 026007 (2010)
  [arXiv:0909.2639 [hep-th]].
  
\bibitem{Jensen:2011su} 
  K.~Jensen, S.~Kachru, A.~Karch, J.~Polchinski and E.~Silverstein,
  ``Towards a holographic marginal Fermi liquid,''
  Phys.\ Rev.\ D {\bf 84}, 126002 (2011)
  [arXiv:1105.1772 [hep-th]].

\bibitem{ads2holo} 
 A.~Strominger,
  ``AdS(2) quantum gravity and string theory,''
  JHEP {\bf 9901}, 007 (1999)
  [hep-th/9809027];
  
  M.~Cadoni and S.~Mignemi,
  ``Asymptotic symmetries of AdS(2) and conformal group in d = 1,''
  Nucl.\ Phys.\ B {\bf 557}, 165 (1999)
  [hep-th/9902040];
  
  J.~Navarro-Salas and P.~Navarro,
  ``AdS(2) / CFT(1) correspondence and near extremal black hole entropy,''
  Nucl.\ Phys.\ B {\bf 579}, 250 (2000)
  [hep-th/9910076];
  
   T.~Hartman and A.~Strominger,
  ``Central Charge for AdS(2) Quantum Gravity,''
  JHEP {\bf 0904}, 026 (2009)
  [arXiv:0803.3621 [hep-th]];
  
  D.~Grumiller, M.~Leston and D.~Vassilevich,
  ``Anti-de Sitter holography for gravity and higher spin theories in two dimensions,''
  arXiv:1311.7413 [hep-th].
  
\bibitem{Preskill:1991tb} 
  J.~Preskill, P.~Schwarz, A.~D.~Shapere, S.~Trivedi and F.~Wilczek,
  ``Limitations on the statistical description of black holes,''
  Mod.\ Phys.\ Lett.\ A {\bf 6}, 2353 (1991).
  
\bibitem{Balasubramanian:1999re}
  V.~Balasubramanian and P.~Kraus,
  ``A Stress tensor for Anti-de Sitter gravity,''
  Commun.\ Math.\ Phys.\  {\bf 208} (1999) 413
  [hep-th/9902121].
  
\bibitem{Spradlin:1999bn} 
  M.~Spradlin and A.~Strominger,
  ``Vacuum states for AdS(2) black holes,''
  JHEP {\bf 9911}, 021 (1999)
  [hep-th/9904143].
  
  \bibitem{loops}
  A.~Almheiri and J.~Polchinski,
  Work in progress.

\bibitem{Banks:1992ba} 
  T.~Banks, A.~Dabholkar, M.~R.~Douglas and M.~O'Loughlin,
  ``Are horned particles the climax of Hawking evaporation?,''
  Phys.\ Rev.\ D {\bf 45}, 3607 (1992)
  [hep-th/9201061].
  
\bibitem{Giddings:1992ff} 
  S.~B.~Giddings and W.~M.~Nelson,
  ``Quantum emission from two-dimensional black holes,''
  Phys.\ Rev.\ D {\bf 46}, 2486 (1992)
  [hep-th/9204072].
 
\bibitem{Almheiri:2013wka} 
  A.~Almheiri and J.~Sully,
  ``An Uneventful Horizon in Two Dimensions,''
  arXiv:1307.8149 [hep-th].
  
  \bibitem{Almheiri:2013hfa} 
  A.~Almheiri, D.~Marolf, J.~Polchinski, D.~Stanford and J.~Sully,
  ``An Apologia for Firewalls,''
  JHEP {\bf 1309}, 018 (2013)
  [arXiv:1304.6483 [hep-th]].
 \end{thebibliography}

\end{document}